\documentstyle[aaspp]{article}
\slugcomment{Astrophysical Journal}

\def\Ha    {H$\alpha$} 
 
\def\NIIb  {[N${\scriptstyle\rm II}$]$\lambda$6548}
\def\NIIr  {[N${\scriptstyle\rm II}$]$\lambda$6583}

\def\NII   {[N${\scriptstyle\rm II}$]}
\def\OI    {[O${\scriptstyle\rm I}$]}
\def\OII   {[O${\scriptstyle\rm II}$]}
\def\OIII  {[O${\scriptstyle\rm III}$]}
\def\SII   {[S${\scriptstyle\rm II}$]}
\def\SiII  {[Si${\scriptstyle\rm II}$]}
\def\CII   {[C${\scriptstyle\rm II}$]}
\def\HeI   {He${\scriptstyle\rm ~I~}$}

\def\Hzero  {H$^{\scriptscriptstyle o}$}
\def\Nzero  {N$^{\scriptscriptstyle o}$}
\def\Ozero  {O$^{\scriptscriptstyle o}$}
\def\Hplus  {H$^{\scriptscriptstyle +}$}
\def\Nplus  {N$^{\scriptscriptstyle +}$}
\def\Oplus  {O$^{\scriptscriptstyle +}$}

\def\Npplus {N$^{\scriptscriptstyle ++}$}
\def\Opplus {O$^{\scriptscriptstyle ++}$}

\def\tauLL {\tau_{\scriptscriptstyle LL}}

\def\deg   {$^\circ$}
\def\kms   {\ km s$^{-1}$}

\def\Msun  {$M_\odot$}
\def\muB   {$\mu_B$}

\def\eg    {{\it e.g.,\ }}

\def\etal  {{\it\ et al.}}
\def\intensity{\ifmmode{{\rm\ erg\ cm}^{-2}{\rm\ s}^{-1}
      {\rm\ Hz}^{-1}{\rm\ sr}^{-1}}
      \else {\ erg cm$^{-2}$ s$^{-1}$ Hz$^{-1}$ sr$^{-1}$}\fi}

\def\flux{\ifmmode{{\rm erg\ cm}^{-2}{\rm\ s}^{-1}}\else {erg
cm$^{-2}$ s$^{-1}$}\fi}

\def\fluxdensity{\ifmmode{{\rm erg\ cm^{-2}\ s^{-1}\ Hz^{-1}}}\else {erg
cm$^{-2}$ s$^{-1}$ Hz$^{-1}$}\fi}
\def\phoflux{\ifmmode{{\rm\ phot\ cm}^{-2}{\rm\ s}^{-1}}\else {\ phot
cm$^{-2}$ s$^{-1}$}\fi}
\def\eflux{\ifmmode{{\rm\ erg\ cm}^{-2}{\rm\ s}^{-1}}\else {\ erg
cm$^{-2}$ s$^{-1}$}\fi}

\def\nH {\ifmmode{\rm n_{\scriptscriptstyle H}}\else{n$_{\scriptscriptstyle H}$}\fi}
\def\NH {\ifmmode{{\rm N_{\scriptscriptstyle H}}}\else{N$_{\scriptscriptstyle H}$}\fi}
\def\Np {\ifmmode{{\rm N_{\scriptscriptstyle p}}}\else{N$_{\scriptscriptstyle p}$}\fi}

\def\ne {\ifmmode{\rm n_{\scriptscriptstyle e}}\else{n$_{\scriptscriptstyle e}$}\fi}
\def\np {\ifmmode{\rm n_{\scriptscriptstyle p}}\else{n$_{\scriptscriptstyle p}$}\fi}

\def\nHo {\ifmmode{\rm n_{\scriptscriptstyle H}^o}\else{$n_{\scriptscriptstyle H}^o$}\fi}
\def\neo {\ifmmode{\rm n_{\scriptscriptstyle e}^o}\else{$n_{\scriptscriptstyle e}^o$}\fi}

\def\RH {\ifmmode{\rm R_{\scriptscriptstyle H}}\else{R$_{\scriptscriptstyle H}$}\fi}
\def\Redge {\ifmmode{\rm R_{\scriptscriptstyle 25}}\else{R$_{\scriptscriptstyle 25}$}\fi}
\def\Fkpc  {\ifmmode{\rm F_{\scriptscriptstyle kpc}}\else{F$_{\scriptscriptstyle\rm kpc}$}\fi}

\def\emunits {\ifmmode{\rm\ cm^{-6}\ pc}\else{\ cm$^{-6}$\ pc}\fi}
\def\Em      {\ifmmode{{\cal E}_m}\else{${\cal E}_m$}\fi}
\def\aB      {\ifmmode{\alpha_{\scriptscriptstyle\rm B}}\else{$\alpha_{\scriptscriptstyle\rm B}$}\fi}

\def\nuo   {\ifmmode{\nu_o}\else{$\nu_o$}\fi}
\def\nuOB  {\ifmmode{\nu_{\scriptscriptstyle\rm OB}}\else{$\nu_{\scriptscriptstyle\rm OB}$}\fi}

\def\phiI  {\ifmmode{\varphi_i}\else{$\varphi_i$}\fi}
\def\phiIV {\ifmmode{\varphi_4}\else{$\varphi_4$}\fi}
\def\Pcos  {\ifmmode{\Phi^o}   \else{$\Phi^o$}\fi}
\def\Jtwo  {\ifmmode{J_{-21}}  \else{$J_{-21}$}\fi}
\def\Jcos  {\ifmmode{J_{-21}^o}\else{$J_{-21}^o$}\fi}

\def\sigHI {\ifmmode{\sigma_{\scriptscriptstyle\rm HI}}\else{$\sigma_{\scriptscriptstyle\rm HI}$}\fi}
\def\sigFP {\ifmmode{\sigma_{\scriptscriptstyle\rm FP}}\else{$\sigma_{\scriptscriptstyle\rm FP}$}\fi}
\def\vmax  {\ifmmode{v_{\scriptscriptstyle\rm max}}\else{$v_{\scriptscriptstyle\rm max}$}\fi}

\begin{document}

\title{Where do the disks of spiral galaxies end?}

\author{J. Bland-Hawthorn} \affil{Anglo-Australian Observatory, P.O.
  Box 296, Epping, NSW 2121, Australia}

\author{K.C. Freeman \& P.J. Quinn\footnote{{\it Current Address:}\
  European Southern Observatory, Karl-Schwarzschild-Strasse 2, 85748 Garching bei M\"{u}nchen, Germany}} \affil{Mt. Stromlo \& Siding
  Springs Observatories, The Australian National University, Private
  Bag, Weston Creek P.O., ACT 2611, Australia}

\begin{abstract}
  In spiral galaxies, the HI surface density declines with increasing
  radius to a point where it is seen to truncate dramatically in the
  best observed cases. It was anticipated that if the ambient
  radiation field is sufficiently strong, there exists a maximum
  radius beyond which the cold gas is unable to support itself against
  ionization. We have now succeeded in detecting ionized gas beyond
  the observed HI disk in spirals. Here, we report on our findings for
  the Sculptor galaxy NGC 253. The HI disks in Sculptor galaxies
  extend to only about 1.2~\Redge\ although we have detected ionized
  gas to the limits of our survey out to 1.4~\Redge. This has important 
  ramifications for spiral galaxies in that it now
  becomes possible to trace the gravitational potential beyond where
  the HI disk ends. The detections confirm that the rotation curve
  continues to rise in NGC 253, as it appears to do for other Sculptor
  galaxies from the HI measurements, but there is a hint that the rotation
  curve may fall abruptly not far beyond the edge of the HI disk. If this is
  correct, then it suggests that the dark halo of NGC 253 may be truncated
  near the HI edge, and provides further support for the link between
  dark matter and HI. The line ratios are anomalous with \NIIb\
  to H$\alpha$ ratios close to unity.  While metallicities at these large 
  radii are uncertain, such enhanced ratios compared to solar-abundance HII 
  regions (\NIIb/\Ha\ $=0.05-0.2$) are likely to require 
  selective heating of the electron population without further ionization 
  of N$^{\scriptscriptstyle +}$. We
  discuss the most likely sources of ionization and heating, and the
  possible role of refractory element depletion (e.g. Ca, Si, Fe)
  onto dust grains.
\end{abstract}


\section{\bf Introduction.}
The case for dark matter in galaxies rests primarily on those spiral
galaxies where the HI rotation curve has been measured at radii which
are several times larger than the optical disk. At these large radii,
the observed rotation of the gas is fully a factor of two larger than
that expected for circular orbits in the potential field of the
luminous galaxy (van Albada\etal\ 1985). Many of these extensive
rotation curves are flat or even rising at the outermost points which
means the edge of the mass distribution has not yet been found (Puche
\& Carignan 1991). In an attempt to find the edge, several observers
have pushed their HI observations to higher sensitivity (van Gorkom
1993; Corbelli\etal\ 1989).  These observations have
shown that HI disks are abruptly truncated at column densities near
10$^{19}$ atoms cm$^{-2}$ where the rotation curves are still flat.

One possible explanation for the truncation is that, at large radii,
the thin HI disks become fully ionized by the metagalactic UV
background (Bochkarev \& Sunyaev 1977). If this is the case, the outer
parts of gaseous disks should be emitting H$\alpha$ photons. Maloney
(1993) has shown that the expected H$\alpha$ emission measure from the
ionized disk beyond the truncation point falls in the range
0.025--0.25 cm$^{-6}$ pc. After demonstrating that these flux levels
are quite feasible with the TAURUS-2 Fabry-Perot interferometer
(Bland-Hawthorn\etal\ 1994, hereafter BTVS), we now use this technique
to search for the signature of ionized gas at and beyond the HI edge
in Sculptor spirals.

The implications of a positive detection are profound.  If the outer
disk could be detected, then we would firstly know how extensive
galactic disks (and dark haloes) really are and, secondly, we could
continue tracking the rotation curve in order to find the edge of the
halo and hence find the total mass of galaxies. A positive detection
also has important implications for explaining low redshift Ly$\alpha$
absorption-line systems towards quasars, and for constraining the
poorly known metagalactic ionizing background.

In $\S$2, we describe the observations carried out at the AAT.
In $\S$3, we derive the expected level of \Ha\ emission at an HI
edge, and then describe the experimental procedure to achieve such
faint detections. The reduction and analysis steps are briefly 
outlined in $\S$4, the results of which are discussed in $\S$5.
In $\S$6, we attempt a physical interpretation of both the gas 
kinematics and the gas excitation before drawing our conclusions in 
$\S$7.

\section{\bf Observations.} 
The observations were carried out over three long dark nights (1994
Aug 9$-$12) at the f/7.91 Cassegrain focus of the AAT 3.9m. Follow-up
observations were carried out at f/14.9 on 1995 Sep 27.  The TAURUS-2
interferometer was used in conjunction with the refurbished HIFI
40$\mu$m gap Queensgate etalon and a 4$-$cavity blocking filter (90\%
peak transmission), centered at $\lambda$6555\AA, with a 45\AA\ 
bandpass well matched to the etalon free spectral range (54.3\AA\ at
H$\alpha$). The 50mm filter was placed out of focus close to the focal
plane and baffled to give a 5.0\arcmin\ field.  The etalon was tilted
by 3.4\deg\ to shift the optical axis to the field edge. An in-focus,
focal plane, colander mask was used to ensure that low-order ghosts
fall outside the field of view. The TAURUS-2 f/8 pupil diameter is
59.9mm which is oversized for the the 50mm diameter etalon. Due to
uncertainties of the precise location of the optical cavity within the
etalon, we placed a 45.0mm aperture stop immediately in front of the
etalon. The pupil stop introduced a major loss of light (50\%)
compounded by a Cassegrain hole which is 17\% of the total pupil area.
The full pupil was used in the f/15 observations. The observational
set-up was the same except for a 75mm blocking filter centered at
6585\AA\ which was not baffled thereby producing a field of view
similar to the f/8 observations.

Observations were made at four positions in NGC 253 (Plates 1 and 2)
with several discrete tunings of the etalon at each position (Table
1). The etalon was used at fixed gap spacings and the resulting ring
pattern was imaged onto a Tek 1024$^2$ CCD with pixel scales of
0.594\arcsec\ pix$^{-1}$ (f/8) and 0.315\arcsec\ pix$^{-1}$ (f/15) with
read noise $\approx$ 2.3e$^{-}$. The wavelength range
$\lambda\lambda 6550-6590$ is dispersed quadratically over the
5\arcmin\ field at 1\AA\ resolution with blue wavelengths to the
north. The instrument and detector were rotated to a position angle of
230\deg\ to align the detector with the galaxy major axis.  At both
foci, we observed two stellar flux standards (Table 1); at f/8, we
also observed four planetary nebulae. Observations were also made of
blank fields and in the direction of the Smith high velocity cloud
(Smith 1963; Wakker 1991), and twilight flats were taken on all nights
with and without the etalon.

\section{Experiment}
\subsection{The expected \Ha\ flux}
We now derive the expected flux levels from spiral edges ionized by a
metagalactic radiation field.  The surface density of HI gas falls
exponentially in spiral galaxies.  A point is reached where the HI
column is no longer able to support itself against ionization by the
cosmic radiation field. A na\"{i}ve calculation can be used to predict
the expected column density \NH\ at which this is likely to occur, in
addition to the local electron density, \ne, and the expected
emission measure, \Em. More sophisticated treatments are given in
Maloney (1993) and Dove \& Shull (1994).

We shall assume that, at large galactocentric radius, the ionization
rate is in rough equilibrium with the rate of recombination.
Therefore, along an axis $z$ perpendicular to the galactic disk, we find
\begin{equation}
  \label{ion_balance}
  2\aB\ \int_0^{\infty} \ne\ \np\ {\rm d}z = \int_{\nuo}^{\infty}\ {{4\pi J_\nu}\over{h\nu}}\ {\rm d}\nu
\end{equation}
where the recombination coefficient \aB\ depends on the neutral
hydrogen column and the local electron temperature (Case B: $\aB
\approx 2\times 10^{-13}$ cm$^3$ s$^{-1}$ at T$_e = 10^4$K).  The
local electron density \ne\ (and therefore the proton density \np), is
related to the neutral hydrogen density, \nH, through the ionizing
fraction $\chi$ such that $\ne = \chi \nH$. Equation~\ref{ion_balance}
assumes that the neutral gas sheet is ionized from both sides. We
approximate the cosmic ionizing flux level as
\begin{equation}
J_\nu = 10^{-21} \Jcos\ \left(\nuo\over\nu\right)^\beta\ \intensity
\end{equation}
where \Jcos\ is the metagalactic flux at the Lyman limit ($\nu =
\nuo$) in units of \intensity. If the cosmic ionizing field (e.g.
Haardt \& Madau 1996) is due largely to quasars at high redshift,
$\beta$ is approximately unity.

For an exponential disk with density profile $\nH\ = \nHo\ 
\exp(-z/z_s)$, the electron density in the plane of the galaxy at the
ionized edge is
\begin{equation}
  \label{elec_dens}
  \nHo \approx 1.8\ \chi^{-1} \sqrt{\Jcos\ (z_s\ \beta)^{-1}}\ \ {\rm cm^{-3}}
\end{equation}
where the vertical scale height of the disk $z_s$ is in parsecs. We
assume a value of $z_s = 100$ pc throughout.  The particle
column density perpendicular to the disk is
\begin{eqnarray}
  \label{col_dens}
  \NH &=& 2 \int_0^\infty \nH\ {\rm d}z \\
      &\approx& 1.1\times 10^{19}\ \chi^{-1}\ \sqrt{\Jcos\ z_s\ \beta^{-1}}\ \ {\rm cm^{-2}}.
\end{eqnarray}
For a $2\sigma$ upper limit of $\Jcos = 0.08\ \intensity$ (Vogel\etal\ 
1995; Bland-Hawthorn 1997), the gas is fully ionized when $\nHo\ 
\approx 0.05$ cm$^{-3}$ and $\NH \approx 3\times 10^{19}$ cm$^{-2}$.
The latter is in good agreement with observations (e.g. Corbelli\etal\ 
1989).  A crude upper limit on the \Ha\ emission measure \Em\ is
\begin{equation}
  \label{em}
  \Em = \int \ne\np\ {\rm dl}\ \ \approx\ 0.25\ \emunits .
\end{equation}
This surface brightness, equivalent to 90 milliRayleighs (mR) and
$5\times 10^{-19}$ erg cm$^{-2}$ s$^{-1}$ arcsec$^{-2}$, is reached at
the 3$\sigma$ level by the Fabry-Perot `staring' technique in less
than one hour of observation (BTVS).\footnote{1 Rayleigh is $10^6/4\pi$
  phot cm$^{-2}$ s$^{-1}$ sr$^{-1}$ or 2.41$\times$10$^{-7}$ erg
  cm$^{-2}$ s$^{-1}$ sr$^{-1}$ at \Ha.}

\subsection{Fabry-Perot `staring'}
We exploit the `staring' technique to obtain a single,
extremely deep spectrum of a diffuse source which fills the field of
view. For a fixed gap spacing, $\lambda \propto \cos\theta$ where
$\theta$ is the angle of an incoming ray of wavelength $\lambda$ to
the optical axis.  The spectrum in a narrow band ($\sim$40\AA) is
dispersed radially from the optical axis across the field (see Fig.
3). Like long-slit spectrometers, the instrumental profile is
projected onto the detector but the line FWHM in pixels varies across
the field inversely with $\theta$. Complete Fabry-Perot rings have
constant surface brightness and equal spectroscopic resolution.  After
flatfielding and point source removal, the data are binned azimuthally
and resampled to a linear axis to obtain a single deep spectrum (e.g.
Figs. 4 and 5).

Maximum sensitivity is achieved when (a) the detector covers the
unvignetted field, (b) the line-emitting source fills the field of
view, and (c) the velocity range and internal kinematic dispersion,
taken together, are comparable to the instrumental profile width.
These criteria determine which galaxies are best matched to a given
instrument.

The isovelocity contours of a disk undergoing flat rotation show two
distinct regimes. The central region of solid body rotation gives rise
to contours which are parallel to the kinematic minor axis.  At the
turnover radius, the contours become radial and therefore subtend a
fixed angle as seen on the sky about the kinematic major axis.  The
spread in velocity within a fixed field of view declines inversely
with radius until we reach the intrinsic spread of an isothermal HI
disk (\sigHI $\approx$ 10\kms; Kamphuis 1993).  An important concept
is the angular extent of the ``innermost monochromatic field'' (imf)
for a given instrument, telescope and object. We define this as the
field at the smallest radius along the disk major axis within which
the observed spread in velocities does not exceed the instrumental
resolution. The necessary condition for a galaxy with maximum
deprojected velocity \vmax\ inclined at an angle $i$ is
\begin{equation}
  \label{imf}
\frac{1}{4} v^2_{\rm max} \sin^2 i
\left[1+\left({{\mu F_{\rm kpc}}\over{2 R \sin i}}\right)^2\right]^{-1}
+ \sigma^2_{\rm HI}\ \  \leq\ \  \sigma^2_{\rm FP}
\end{equation}
where \sigFP\ is the instrumental width (in \kms) and $R$ is the
radius to the field center (in kpc). Truncated HI edges typically
occur at radii larger than the semi-major axis distance (\Redge\ in
kpc) of the $B = 25$ mag arcsec$^{-2}$ contour. For a fixed field
\Fkpc\ (in kpc), this condition is most easily satisfied in large,
nearby galaxies where the instrument field of view is a small fraction
of the object size. The constant $\mu$ allows for two different
approaches to sky subtraction.  If this is achieved through a separate
exposure, $\mu = 1$; otherwise, in a properly matched experiment, we
assume that half the field is given over to the sky background for
which $\mu = 0.5$.

The optimal disk inclination is unclear: face-on galaxies have a
smaller velocity spread and subtend larger solid angles, whereas an
inclined disk has a higher projected column density. A potential
problem in edge-on disks is flaring due to the declining stellar
surface density (Olling 1995) which tends to reduce the projected
emission measure.  To minimize the velocity spread, we rotate the
instrument so that the incomplete ring pattern is convex in the same
sense as the projected velocity field, and is centered on the
kinematic major axis (see Figs. 1 and 2).

\section{Reduction \& Analysis}
A detailed discussion of the many subtleties of data analysis are to
be presented elsewhere; a brief overview is given here. An important
step is establishing the optical axis of the incomplete calibration
rings to better than 0.1 pix.  This was achieved with orthogonal
distance regression (e.g. Boggs, Byrd \& Schnabel 1987). Next, we
azimuthally bin the calibration rings for all nights to ensure that
(a) there were no opto-mechanical shifts, (b) the instrumental
response was constant, and (c) the etalon gap zero-point was constant
from night to night. In (b), the etalon was found to behave reliably
except that there was a slow drift in the optical gap during the first
night.

Some calibration spectra showed baseline variations after binning over
different parts of the field. This arises from illumination (vignette)
effects, stray light, chip structure and CCD fringeing which can be
divided out reliably with flatfields.  The spectral bandpass seen by
individual pixels is roughly 1\AA\ where the bandpass centroid
declines by 40\AA\ from the optical axis to the field edge. Twilight
flats were found to be the most reliable except that the Fraunhofer
spectrum leaves its imprint in the data. We divide out the solar
spectrum from the flatfield by establishing the mean spectrum and then
generating a polar image with this spectrum. Occasionally, the
flatfields leave residual fringe structure in the data. It is possible
to remove this with azimuthal smoothing but potentially informative,
intensity variations in the data will be washed out. CCD fringeing
constitutes the main systematic error in diffuse detection and
provides a major challenge to achieving deeper detection limits than
that quoted by BTVS.  The response of the blocking filter is removed
in the twilight division.

To obtain discrete spectra from the summed CCD images, the data were
divided into annular rings two pixels wide. Within each annulus, we
can calculate the mean, median or mode of the histogram. Cosmic ray
events appear as outliers and are therefore easily removed.  We tried
more sophisticated methods (e.g. the bi-weight statistic) but the
improvement was found to be marginal.  The weak underlying galaxy
continuum observed in NGC 253 had only a negligible effect on the
final spectra. This was examined by subtracting a matching $R$ band
exposure modulated by the polar response of the blocking filter.  This
same image was used to identify unresolved sources within the field.

Sky subtraction is particularly hazardous since atmospheric humidity
produces variable water vapour features. If the object fills the field
of view, as for NGC 253, it is necessary to obtain off-field exposures
at the same airmass and at comparable humidity.  The OH lines are time
variable but, most importantly, they vary with the zenith angle of the
observation (Kondratyev 1969). This is quite distinct from the
geocoronal lines which reach a minimum at local midnight rising
sharply towards dawn. The night sky lines allow for wavelength
calibration to better than 0.04\AA\ (Osterbrock \& Martel 1992) as
shown by the difference in wavelength produced by a discharge lamp and
solar Ly$\beta$ resonance excitation of the exosphere (Yelle \&
Roesler 1985). The velocity centroid of the galactic \NII\ emission
(`Reynolds layer'), using new wavelengths from Spyromilio (1995),
provides an independent measurement of the Earth-Sun motion with
respect to the Local Standard of Rest. For all four nights, the
conditions were consistently photometric as judged by the Reynolds
layer emission. The photometric calibration was achieved using both
line and continuum flux standards (Table 1).

\section{Results}

The four TAURUS-2 field positions are shown in Figs. 1 and 2 overlaid
on a $B$ image of NGC 253 and the deepest HI image to date for this
galaxy. A magnified section of one interferogram is shown in Fig. 3
and summed spectra are shown for the two outermost fields in Figs. 4
and 5.  The `staring' method allows for an extremely deep
spectroscopic detection over pixels that fall within the projected arc
defined by a discrete frequency. There are two frequencies of interest
so that kinematic measurements are possible at two positions within
each field. The etalon was tuned to two different spacings for the two
innermost fields so that we have kinematic measurements at no more
than 12 independent positions along the major axis. These data are
presented in Figs. 6 and 7.

Optical line emission is detected in all field positions. In fields
SW1 and SW2, bright HII regions were clearly seen in both \Ha\ and
\NIIb.  These short exposures were taken to establish
independently the systemic velocity of the gas. The kinematic
measurements are seen in Fig. 6$c$ at $r < 7$\arcmin. The large
systematic errors in both axes are due to azimuthally averaging over a
large spread in velocities and radial distances within the projected
plane at small galaxian radii.

Field SW3 is the imf defined by the criterion in equation~\ref{imf}.
Weak stellar continuum emission is seen over the central third of the
field extending from roughly \muB\ $=$ 23 mag arcsec$^{-2}$ (NE) to 25
mag arcsec$^{-2}$ (SW). The \NII\ and \Ha\ line emission comprise
clumpy structure superimposed on a faint diffuse component. Only two
of the sources are unresolved: these show up on our matched $BVR$
images and are labelled in Fig. 3. The two bright knots at the
position of the \NII\ line are probably faint HII regions.

The summed spectrum for field SW3 is shown Fig. 4. The difference of
the on-object and off-object spectra clearly shows the declining
stellar disk along the major axis, in addition to the \NII\ and \Ha\ 
lines at 6557.7\AA\ and 6572.8\AA. While these lines fall at different
spatial locations within the galaxy, the strength of the blue \NII\ 
line with respect to \Ha\ is striking, a result which holds after
removing the bright knots. The weakness of the galaxy continuum rules
out a significant correction to the \Ha\ line flux due to stellar
absorption (Bica \& Alloin 1987). We achieved the same basic result in
the 1995 September 27 observations although the signal-to-noise ratio
was roughly half that of the earlier observations. Both sets of
observations show that the diffuse emission peaks towards the major
axis.

The summed spectrum for field SW4 is presented in Fig. 5. This field
lies beyond the HI edge of NGC 253.  Unfortunately, we did not manage
to obtain a matching sky observation. The data clearly show weak \NII\
emission and possibly broad diffuse \Ha\ emission. We have attempted
to subtract something approximating a background spectrum by dividing
up the field into three vertical panels which have equal area when the
outer panels are taken together. The central section constitutes our
on-object spectrum, the outer quarter panels our off-object spectrum.
The weak outer envelope of the galaxy is seen in the difference
spectrum. The \NII\ emission peaks towards the major axis and is not
detected in the outermost panels. Broad diffuse \Ha\ emission is seen
in both regions at comparable intensity and therefore subtracts
cleanly in the residual spectrum. That this feature is real can be
seen by comparing the lower spectrum in Fig. 5 with the off-object
spectrum in Fig. 4 where such a feature is not seen.

The radial velocities for the emission lines in Figs. 4 and 5 were
corrected for the Earth's motion before subtracting the systemic
velocity of NGC 253. These measurements are presented in Figs. 6 and
7.  The innermost TAURUS-2 measurements were used to establish the
systemic velocity of the ionized gas. Schommer\etal\ (1993) show the
systematic uncertainties which can arise between HI and ionized gas
kinematics. Our optical determination ($243\pm 9$ \kms) is consistent
with the weighted-average HI measurements (249$\pm$8\kms) from the
NASA/IPAC Extragalactic Database. We adopt the HI value 245$\pm$5\kms\ 
(Puche\etal\ 1991, hereafter PCvG) as we wish to relate the optical
data to their data.  For the measurements between $r=10$\arcmin\ and
$r=14\arcmin$, the kinematic error is dominated by the uncertainty in
systemic velocity.  The smaller error bars compared to the inner
TAURUS-2 measurements reflect the smaller intrinsic radial and
kinematic dispersion.  That the optical values appear systematically
high by 10\kms\ is discussed in the next section.  For the outermost
point, the diffuse broad feature has a large kinematic measurement
uncertainty.  Unlike the other lines, this feature has fairly uniform
intensity over the full field.  Its lack of continuity with the other
measurements calls into question its association with the galaxy.

\section{Discussion}

\subsection{Dynamical interpretation}
The major axis measurements in NGC 253 pose an interesting problem. In
Fig. 6, the ``rotation curve'' appears to flatten off before rising at
a radius $r=7$\arcmin. This effect is seen for both the approaching and
receding sides of the disk, with (Fig. 6$b$) or without (Fig. 6$a$)
the correction for orbit inclination. In Fig. 6$c$, the TAURUS-2
measurements confirm and extend this trend out to $r=13\farcm 5$.
The outer TAURUS-2 values are somewhat higher than the azimuthally
averaged HI rotation curve, but this is a distinctive feature of the 
approaching side of the disk (Fig. 6$a$) in that the outer
HI measurements here are higher than average.

If we assume that the distribution of mass has spherical symmetry,
then from Poisson's equation,
\begin{equation}
{{d}\over{dr}}(r v_c^2) = 4\pi G r^2 \rho(r)
\end{equation}
where $\rho$ and $v_c$ are the local mass density and the circular
velocity. For the flat region of $v_c$, $\rho$ declines as $r^{-2}$;
the sudden rise requires the density profile to become much
flatter or even constant with increasing radius. A slightly flattened
potential gives the same result although the inferred total mass will
be overestimated in this case.

The inflexion in the rotation curve can be understood naturally to
arise from the potential of an exponential stellar disk with a much
larger dark halo (e.g. Carignan \& Freeman 1985).  To illustrate this,
we show a variety of possible 3-component fits to the data.

In the first model (Fig. 7$a$), the HI and stellar disks have the same
form. The dark halo is represented by an isothermal sphere, which is
fully specified by any two of the halo central density, $\rho_h$, the
central velocity dispersion $\sigma_h$, or the core radius $r_h$. The
stellar disk is specified by a central surface density $\mu_e$ and a
scale length $r_e$.  The HI surface density profile is given in PCvG
(Fig. 5); the central hole within $r=1\farcm 9$ causes the
gravitational force due to the gas to be directed outward (e.g.
Staveley-Smith\etal\ 1990). We have taken $\sigma_h$ $=$ 253\kms, $r_h
= 18$ kpc, $\mu_e = 1.35\times 10^9$\ \Msun\ kpc$^{-2}$ and $r_e = 1.6$
kpc. These values compare well with PCvG except that the disk and halo
scale lengths are 30\% smaller than their values.  In order to match
the observed data, the model rotation curves have been projected to a
plane inclined at 72\deg. We note that such a large core radius is
normal for the dark halo of large disk galaxies (\eg Freeman 1993).

In the second model (Fig.  7$b$), the radial forms of the disk and
halo are given by Dehnen \& Binney (1997). For the halo, we adopt
\begin{equation}
  \label{dehnen1}
  \rho_h(R,z) = \rho_o \left(m\over r_o\right)^{-g} \left(1+{m\over
  r_o}\right)^{g-b} \exp(-m^2/r_t^2)
\end{equation}
with
\begin{equation}
  \label{dehnen2}
  m^2 = R^2 + (z/q)^2.
\end{equation}
The halo parameters are the halo density normalization, $\rho_o$, the
halo scale radius, $r_o$, the outer cut-off radius, $r_t$, and the
inner and outer power slopes $g$ and $b$. Apart from the halo axis
ratio ($q=0.8$), all of these parameters were obtained from a fit to
the rotation curve.  Both the stellar and gaseous disks are assumed to
be exponential in form, i.e.
\begin{equation}
  \label{dehnen3}
  \rho_d(R,z) = {{\mu_d}\over{2 z_d}}
  \exp\left(-{{R_m}\over{R}}-{{R}\over{R_d}} - {{\vert z \vert}\over{z_d}}\right).
\end{equation}
For each of the disks, the adopted parameters are the surface density
normalization, $\mu_d$, the disk scale height, $z_d$, and the disk
scale radius and inner cut-off radius, $R_d$ and $R_m$. For the HI
disk, these were fixed at $\mu_d = 1.94\times 10^7$\ \Msun\ 
kpc$^{-2}$, $R_d = 3.92$ kpc, $z_d = 0.04$ kpc, and $R_m = 1.22$ kpc.
For the stellar disk, $R_d = 1.80$ kpc, $z_d = 0.15$ kpc, and $R_m =
0$ kpc as suggested by the photometry. The surface density
normalization for the stellar disk was obtained from the least-squares
fit. 

The fit, kindly undertaken by W. Dehnen, ignored the last TAURUS-2
measurement to yield the following results: $\mu_d = 1.56\times
10^9$\ \Msun\ kpc$^{-2}$, $\rho_o = 1.98\times 10^8$\ \Msun\ 
kpc$^{-3}$, $g = -1.058$, $b=1$, $r_o = 34.2$ kpc and $r_t = 164.0$
kpc.  Of these, $b$ was constrained to be greater than or equal to
unity.  Fig.  7$b$ shows that the rising halo contribution to the
rotation curve gives a better fit to the data overall. But there are
strong correlations between the various parameters and the fitted
results depend critically on which of the TAURUS-2 points are
included.

The outermost TAURUS-2 \Ha\ measurement in Figs. 6$c$ and 7 suggests
that the rotation curve may be falling beyond a radius of about 10
kpc, but there are certainly other possible explanations for its low
observed velocity.  For example, {\it (i)}\ it may result from tidal
distortion of the outermost regions of NGC 253, or {\it (ii)}\ from
accretion of a faint gas-rich object at large radius. In support of
the accretion idea, a stacked photographic image by D. Malin shows
extended stellar light at faint levels ($\sim$26 mag arcsec$^{-2}$) to
the SW of the disk; similar emission is not seen to the NE. {\it
  (iii)}\ While the emission may arise from an outer ionized envelope,
it is conceivable that this emission is associated with the group
itself. An extragalactic \Hplus\ cloud has recently been discovered in
the Fornax cluster (Bland-Hawthorn\etal\ 1995). Mathewson, Cleary \&
Murray (1975) and Arp (1985) found tentative evidence for HI clouds
within the Sculptor group (cf.  Haynes \& Roberts 1979). It is
difficult, however, to reconcile the \Ha\ radial velocity
($\approx$440\kms) with the Magellanic Stream which passes in front of
the Sculptor group (Mathewson \& Ford 1984).

Finally, we consider the possibility that the rotation curve of NGC
253 is indeed falling beyond 10 kpc. Such a rapid decrease suggests
that the dark halo may be truncated. We can illustrate the effect of
such truncation with a simple model.  We take the halo density
distribution to be
\begin{equation}
\rho_h = \rho_\circ (1 + r^2/r_a^2)^{-1}
\end{equation}
with the rotation curve $V(r)$ given by
\begin{equation}
\label{arctan}
V^2 = V_\infty^2 [1 - ({{r_a}\over{r}})\ \tan^{-1} ({{r}\over{r_a}})]
\end{equation}
For $V_\infty = 733$ \kms\ and core radius $r_a = 18$
kpc, the rotation curve for this model is very similar to that of the
isothermal halo in Fig. 7$a$: the optical galaxy lies well within the
core radius of the halo, and the halo rotation curve is close to
solid-body.  We now truncate the halo at $r = 10$ kpc; its rotation
curve is then Keplerian for $r > 10$ kpc. Fig. 7$c$ shows how the
resulting total rotation curve now provides an acceptable fit to all
of the HI and TAURUS-2 data.  The model is artificial (the halo is
sharply truncated, and we did not truncate the stellar and HI disks),
but the point is obvious enough: truncation of the dark halo near the
observed outer edge of the HI disk can produce the apparent falling
rotation curve.

If this is all correct so far, there are some interesting
consequences.  The properties of dark halos are best studied in disk
galaxies for which the HI distribution extends well beyond the optical
distribution. In the outer regions of these galaxies, the ratio of
(dark matter surface density)/(HI surface density) is roughly constant
(\eg Bosma, 1978; Carignan 1991). NGC 253 is not such a galaxy.  Its
HI extent is similar to its optical extent (\eg PCvG). Our data
provides the first hint that, in galaxies like NGC 253 where the HI
and the light are co-extensive, the dark matter also may not extend
much beyond the optical distribution. It is tentative evidence of
the apparent link between the dark matter and the HI. We emphasize
that the dark matter is still essential to generate the observed
rotation curve for NGC 253: the inferred mass ratio of dark matter to
luminous matter is about 5.  Our point is simply that the dark matter
distribution in NGC 253 may be truncated at a radius of only 10 kpc,
compared to the much larger dark halo distributions observed in our
Galaxy (\eg Freeman 1996) and other large spirals (\eg Zaritsky \etal\
1997).

The reduced chi square for the models presented in Fig. 7$a$ and 7$b$
are 5.2 and 2.9 respectively. The latter model has the highest
statistical significance of all models, but the basic assumptions
are unlikely to be physical.  A flat rotation curve given
by equation~\ref{arctan} does not fit the measurements adequately.
The reduced chi square is 6.3 and, more significantly, 10 of 12 points
outside of $r=7$\arcmin\ miss the curve by more than 2$\sigma$. But
the inclusion of a truncation radius lowers the reduced chi square to
3.9 and the curve now passes through all the outer points (Fig. 7$c$).

\subsection{Ionization and heating of the diffuse gas}

The most striking feature of the spectra presented in Figs. 4 and 5 is
the inferred high \NIIb/\Ha\ ratios. In order to ensure that the
whitelight calibration adequately removed the filter response over the
field ($\S$4), we examined the line ratios in HII regions over the
inner disk.  Where we have two interference rings that fall close
together on the sky -- one from \Ha, the other from [NII] -- the HII regions
appear weak in [NII] compared to \Ha, indicative of the order of
magnitude difference expected for inner disk HII regions.  If the
outer HI gas has sub-solar abundances (Pagel 1989; Diaz 1989), the
enhanced line ratio must arise from one or more excitation processes.
However, to date, numerical models involving a single ionizing source
have not managed to produce such an enhancement (e.g.  Sokolowski
1994).  Thus, we separate our discussion of ionization and excitation
wherever possible. Almost certainly, the anomalous ratio is indicative
of a higher local electron temperature for which there are various
mechanisms. In solar-abundance HII regions, this ratio rarely exceeds
0.15 (Evans \& Dopita 1985): to substantially increase this ratio
requires that we selectively heat the electrons without producing
\Npplus.

\nobreak{\it The required ionizing flux.} The emission measure \Em\ 
from the surface of a cloud embedded in a bath of ionizing radiation
gives a direct gauge, independent of distance, of the intensity of the
ambient radiation field beyond the Lyman continuum (Lyc) edge (\eg
Hogan \& Weymann 1987).  This assumes that the covering fraction {\it
  seen by the ionizing photons} is known and that there are sufficient
gas atoms to soak up the incident ionizing photons.  At electron
temperatures of 10$^4$K, collisional ionization processes are
negligible. Perpendicular to the surface, from
equation~\ref{ion_balance}, the column recombination rate in
equilibrium must equal the incident ionizing photon flux, $\aB \ne \Np
= \phiI$, where \phiI\ is the rate at which Lyman continuum photons
arrive at a planar cloud surface (phot cm$^{-2}$ s$^{-1}$), and \Np\ 
is the column density of ionized hydrogen. The emission measure
produced by this ionizing flux is
\begin{equation}
  \label{em_flux}
  \Em\ \approx 4.5 \varphi_4\ {\rm mR}
\end{equation}
where $\phiI = 10^4 \phiIV$ (Bland-Hawthorn \& Maloney 1997, hereafter
BM).  For an optically thin cloud in an isotropic radiation field, the
solid angle from which radiation is received is $\Omega = 4\pi$, while
for one-sided illumination, $\Omega=2\pi$.  However, \phiI\ can be
anisotropic and $\Omega$ can be considerably less than $2\pi$.

\nobreak{\it Metagalactic ionizing background.} The measured surface
brightness values for the emission lines in Figs. 4 and 5 are as
follows: for the \NIIb\ line, 90$\pm$5 mR ($r=10\farcm 5$), 46$\pm$5 mR
($r=12\farcm 0$); for the \Ha\ line, 81$\pm$6 mR ($r=13\farcm 5$),
41$\pm$8 mR ($r=15\farcm 0$). If the underlying HI is optically thick
to the Lyman continuum, the \Ha\ lines require an ionizing flux of
$2\times 10^5$\phoflux\ and $9\times 10^4$\phoflux. The present upper
limit on the metagalactic ionizing flux ($\S$3.1) is \Jcos $<
0.08$\intensity\ ($2\sigma$) which sets an upper limit on the
one-sided ionizing flux at the surface of an optically thick HI sheet
of $2\times 10^4$\phoflux, thereby ruling out the metagalactic
background as the dominant ionizing source. The combined ionizing
radiation from hot gas and galaxies within the Sculptor group is also
much too weak. Sciama (1995) predicts an order of magnitude higher
ionizing flux from decaying neutrinos which could conceivably account
for much of the \Ha\ flux. However, the narrow energy bandpass of
decaying neutrinos (Sciama 1996) at the Lyman limit cannot produce
\Nplus, and therefore requires an additional source of ionization.

\nobreak{\it Compact halo sources.} The most recent summary of
the MACHO project indicates that as much as half of the dark matter in
the Galaxy out to the LMC is made up of solar mass objects. There is
wide disagreement on whether the missing mass could comprise a halo
population of white dwarfs (Adams \& Laughlin 1996; Kawaler 1996;
Chabrier, Segretain \& Mera 1996). But most plausible models invoke a
population which produces essentially no UV flux today. This is
supported by high redshift observations of the precursor halo
population (Charlot \& Silk 1995).

\nobreak{\it Compact disk sources.} Unlike ellipticals, spiral galaxy
disks tend not to ``grow'' when photographic plates are stacked or
amplified to reveal light at very faint levels (Malin 1983). But a
photographically amplified plate of NGC 253 (Malin 1981) reveals a
faint blue disk extending to the limits of our outermost detection
($r=15$\arcmin). Following a suggestion from G.B. Field, we now
consider whether these could comprise an extended white dwarf
population (including the precursor population of central stars in
planetary nebulae) young enough to produce a significant UV disk flux.
Such a population was originally broached by Lyon (1975) and Bania
\& Lyon (1980). In Fig. 8, we consider whether white dwarfs are able
to account for the diffuse line emission in concert with the faint
stellar continuum at the HI edge in NGC 253 (barred arrow). We present
the expected \Ha\ emission measure \Em\ versus $B$ surface brightness
for an isothermal population of white dwarfs as a function of the disk
surface density.  The contribution to both of these quantities from
white dwarfs in the solar neighbourhood is also shown (filled
symbols).  The luminosity function is taken from Liebert, Dahn \&
Monet (1988) and the luminosity$-$temperature conversion from Wood
(1990).

There are two basic problems in using white dwarfs to explain the disk
ionization. In the solar neighbourhood, the combined effect of all
white dwarfs could produce \Em\ $=0.1$\emunits\ which fails to explain
the Reynolds layer emission by an order of magnitude or more, in
agreement with Nordgren, Cordes \& Terzian (1992). If we assume a
comparable star formation history in the outer reaches of NGC 253,
this model could conceivably be used to explain the measured \Ha\ 
emission. This is particularly attractive in that the high
temperatures of the white dwarf population produce elevated \NII/\Ha\ 
ratios (Sokolowski \& Bland-Hawthorn 1991).  However, if the
luminosity function resembles the solar neighbourhood, the combined
disk surface brightness at $B$ from the cooler white dwarfs is very
much brighter than observed, particularly when one includes the
contribution of early main sequence stars. Alternatively, attempting
to explain the low \Ha\ emission measure in terms of 2000 white dwarfs
kpc$^{-2}$ at 45,000 K leads to a timing problem. We have measured an
age-temperature relation from Wood (1990) for which
\begin{equation}
  \label{age-temp}
  \log ({t\over{\rm yrs}}) = 3.6 \log T_{\rm wd} - 9.7.
\end{equation}
Stars hotter than 40,000 K have left the asymptotic giant branch less
than 10$^7$ years ago which would require a highly contrived star
formation history throughout the disk.

S.G. Ryan has suggested the possibility of hot horizontal branch
stars, in particular, an isothermal population with temperatures
around 15,000 K (Lee 1993). If we assume comparable line blanketing
with white dwarfs, Fig. 8 illustrates that such a population will
always produce too much blue light for the required ionizing flux,
irrespective of the uncertain sizes and surface densities of both
objects. The characteristic ``temperature'' of the ionizing source
needs to be sufficiently high to fall within the strict $B$ band
limit. Interestingly, this is just what is needed to significantly
enhance the \NII/\Ha\ ratio as photons with higher mean energies
produce higher electron temperatures.
 
{\it Ram pressure heating.} Mathis (1986) has stressed the problems
associated with shocks as a source of ionization and heating for the
Reynolds layer, particularly the near-uniformity of the emission. An
interesting possibility is ram pressure heating as NGC 253 moves
through an external medium where the disk is inclined at some angle to
its direction of motion. The wake of NGC 7421 (Ryder\etal\ 1996)
suggests that such a process can take place.  To ionize a column of
10$^{19}$ atoms cm$^{-2}$ requires shock speeds close to 130\kms\ 
(Dopita \& Sutherland 1996). A medium with pre-shock density $\sim
10^{-2}$ cm$^{-3}$ moving through an ambient medium with the relatively 
high density $\sim
10^{-3}$ cm$^{-3}$ at 400\kms\ could produce the necessary ram
pressure. The post-shock temperature at a shock velocity of 130\kms\ 
is $2.4\times 10^5 K$ which is a factor of ten higher than the upper
limit from the Doppler parameter (22\kms) of the observed line
profiles. For \Nplus\ and \Hplus\ to have been collisionally ionized,
the gas must have had time to cool.  But groups with low velocity
dispersions, like the Sculptor group, tend to have very little
intracluster medium (Fadda\etal\ 1996). Therefore, the galaxy would
need to be running into a large, external gas cloud for the shock
model to remain plausible.

\nobreak {\it Exotic heat sources.} There exists a wide range of
`exotic' models for preferentially heating the electron population
including turbulence-driven MHD wave heating (Raymond 1990), and
mixing layers driven by bulk flows (Slavin, Shull \& Begelman 1993).
Bulk flows and turbulence in the disk are thought to be maintained
primarily by energy injection from star formation. The star formation
rates at the HI edge are likely to be very small.  An alternative
possibility is that the outer disk is being rained on by galactic
fountain material driven by the inner starburst (e.g.  Benjamin \&
Shapiro 1993). It is unclear at present whether fountain models can
generate the necessary heating through shocks. Ferland \& Mushotsky
(1984) have shown that under rather special conditions, low energy
cosmic ray electrons can selectively heat ions and electrons through Coulomb
repulsion.  However, Sciama (1996) emphasizes the difficulties
involved in cosmic rays penetrating gas clouds and these do not appear
to be an important agent.

\nobreak{\it Young stellar disk.} The inner disk of NGC 253 hosts a
young, stellar population producing copious amounts of ionizing
photons. It is plausible that the outer disk sees this radiation
either through dust scattering or through warps in the outer parts.
For Rayleigh scattering, we have repeated the calculations of Jura
(1980) using up-to-date dielectric phase functions (Martin \& Rouleau
1990; Draine \& Lee 1984) and assuming a standard grain mixture and
distribution (Mathis, Rumpl \& Nordsieck 1977). We adopt the
Henyey-Greenstein (1941) phase function which relates the asymmetry
parameter $g$ to photon energy. The expected ionizing flux at
the HI edge is much less than 10$^5$\phoflux\ due primarily to the
highly forward-scattering behaviour ($g>0.9$) of the grains at
increasing energy towards the Lyman limit.

While there is some uncertainty as to the fraction of ionizing photons
which escape a normal spiral galaxy, two independent lines of argument
suggest that roughly 5\% of ionizing photons escape the Galaxy
(Domg\"{o}rgen \& Mathis 1994; BM). The absolute $B$ magnitude of NGC
253 is comparable to that inferred for the Galaxy and thus, if the
disk opacity is comparable in both cases, the same models in BM apply
here. G.D. Bothun (personal communication) suspects that the disk of
NGC 253 may be more opaque than the Milky Way, in which case our
predicted fluxes should be considered upper limits.  To explain the
\Ha\ flux at both outer positions (9 and 12 kpc) requires an integral
sign warp originating at 7 kpc reaching 20\deg\ at 10 kpc.  It is
difficult to rule this out on the basis of the HI data. At face value,
Fig. 8 of PCvG indicates such a warp starts at 7 kpc reaching
$\approx$10\deg\ at 9 kpc. In the next section, we explore this model
in more detail.

\nobreak {\it Dilute photoionization.} After
Sokolowski (1994), we attempt to simulate conditions at the HI edges
of spiral galaxies ionized by the stellar radiation field from the
central disk.  This requires that the outer SW edge of NGC 253 is
warped ($\sim 25$\deg) or flares up and therefore sees the central
regions. We use the CLOUDY code (Ferland 1991) to ionize plane
parallel slabs with cosmic abundances (Grevesse \& Anders 1989). The
physical state and emission spectrum of a low-density photoionized gas
with a given composition are fixed by two parameters $-$ the shape of
the ionizing continuum and the ionization parameter (Tarter, Tucker \&
Salpeter 1969). For the ionization parameter, $U$ (the ratio of
ionizing photons to nucleons), we adopt
\begin{equation}
  \label{U}
  U = {{\varphi}\over{c \nH}}
\end{equation}
where $c$ is the speed of light. In Fig. 10, the results for a range
of emission-line diagnostics are presented as a function $U$. The
radiation-bounded models are truncated when the electron temperature
$T_e$ falls below 4000K; all models assume \nH $=$ 1 cm$^{-3}$.

The unabsorbed radiation field is a composite taking the form $\psi_c
\xi(M_*) t(M_*)$ where $\psi_c$ is the current star formation rate,
$\xi$ is the initial mass function, and $t$ is the main-sequence
lifetime. The ionizing field is dominated by stars with $M_* \geq 40
M_\odot$; we specify the upper mass cut-off to be $M_* = 120 M_\odot$.
The stellar atmospheres are from Mihalas (1972) and Kurucz (1979); the
solar metallicity evolutionary tracks are from Maeder (1990). For the
massive star IMF, we adopt $\xi(M_*) \propto M_*^{-2.7}$ which falls
between the Salpeter (1955) and the Miller-Scalo (1979) models.

Irrespective of the {\it shape} of the ionizing continuum, a dilute
radiation field leads to enhanced emission from low ionization
emission lines (Ferland \& Netzer 1983; Halpern \& Steiner 1983).  But
a general hardening of the ionizing field also produces the same trend
(e.g. Sokolowski \& Bland-Hawthorn 1991). Indeed, if the edges of
spirals are ionized by the central stellar disk, the ionizing field is
expected to be both dilute and hardened. 

There are at least two phenomena which serve to harden the local
ionizing radiation field: interstellar opacity and metal depletion
(i.e. refractory elements) onto grains.  We assume that the ionizing
photons have leaked from the young inner disk and that the ionizing
spectrum has been hardened by intervening absorption ($\tauLL\sim 3$
perpendicular to the disk).  We adopt the atomic photoionization
cross-section for which the average interstellar opacity varies as
$E^{-2.43}$ above 13.6eV (Cruddace\etal\ 1974). In quiescent galaxies,
the outer HI disk is expected to have sub-solar metallicities (\eg\ 
Molla\etal\ 1996).  But the ongoing nuclear starburst in NGC 253 (Beck
\& Beckwith 1984; Antonucci \& Ulvestad 1988) could conceivably enrich
the outer disk through long-range galactic fountains (Corbelli \&
Salpeter 1988). On balance, we adopt solar abundances (Grevesse \&
Anders 1989) modified by the known depletion rates for cold gas (Cowie
\& Songaila 1986; Jenkins 1987; Savage \& Massa 1987).

\nobreak{\it Gas phase depletion.} The formation of grains depletes 
primarily Ca, Fe and Si which in turn
suppresses the dominant coolants from the singly-ionized stages of
these atoms (Ferland 1992). For example, at $U = 10^{-4}$, \CII
158$\mu$m and \SiII 35$\mu$m account for one quarter of the total
nebular cooling. While C, N and O are not strongly depleted, the
forbidden line emission from \Nplus, \Ozero, \Oplus\ and \Opplus\ is
greatly enhanced by the increased temperature.  In particular, Mathis
(1986) finds that the \NII/\Ha\ ratio increases as $T_e^{3.6}$.
Therefore, a 20\% increase in the local electron temperature can
effectively double this line ratio. We neglect grain heating and
cooling processes (Reynolds \& Cox 1992) as these are only important
in the high $U$ limit (Baldwin\etal\ 1991; Ferland 1992). Complete
removal of the grain population from the depleted gas causes line
strengths to change by no more than 5\% (Shields 1992).

{\it Photoionization models.}
In Fig. 9, we show the detailed ionization and thermal structure
within the gaseous slab. \Nzero, \Ozero\ and \Hzero\ are closely coupled
with slight differences due to charge exchange reactions. \Nplus,
\Oplus\ and \Hplus\ are therefore also closely linked, except
that divergence can occur at the front of the slab if the radiation
field is sufficiently strong to produce a higher state of ionization
in N or O.  The soft photons are soaked up and fully ionize the front
of the slab; the harder photons propagate further into the gas and set
up a partially ionized zone. Here, the electron temperature increases
by 25\% which boosts the emissivities of the \OI, \OII, \NII\ and
\SII\ lines. The high energy photons are essential: cutting off the
continuum just below the ionization potential of N{\scriptsize II}
(29.6 eV) produces \NIIb/\Ha\ line ratios no higher than 0.25.

Fig. 10 shows the dependence of five important line diagnostics on
both ionization parameter and the gas column density.  The range of
column densities shown are relevant to spiral edges. The line ratio
\NIIr/\Ha\ peaks at $U=10^{-4}$ with a fivefold
enhancement when compared with HII regions. In the warped disk model
above, we expect $\phi = 3-30\times 10^4\phoflux$. Only the high end
of this range is sufficient to explain the \Ha\ recombination
emission. For a gas disk with a thickness of 1 kpc, \Em $=$
0.1\emunits\ implies $n_e \approx 0.01$ cm$^{-3}$, such that for a
fully ionized gas, $U = 10^{-5}-10^{-4}$. The high end of this range
is indeed where the \NIIb/\Ha\ ratio peaks in Fig. 10.  
The observed ratio (\NIIb/\Ha\ $\approx$ 1)
implies \NIIr/\Ha\ ($\approx 3$) is a factor of two higher than the
peak of the curve in Fig. 10. While this model may have some
application to the HI edges of NGC 253, a full explanation of the gas
excitation requires an additional heat source.

Our attempt to explain the enhanced \NII/\Ha\ ratios with the dilute,
hardened radiation field is made much more difficult if the particle
column density seen from the nucleus is much less than $3\times 10^{19}$ 
atoms cm$^{-2}$.
It is noteworthy that \OI$\lambda 6300$ is produced deep within the
slab compared to \OII\ and \OIII. The relative strength of these
lines exhibits a complex interdependence on the shape of the ionizing
spectrum, the ionization parameter and, most crucially, the column
density of the ionized gas. Fig. 10 illustrates that, in fact, all
emission lines exhibit some dependence on the gas column density. 

Rand (1997) emphasizes the importance of the \HeI$\lambda 5876$/\Ha\ 
and the \NII/\Ha\ ratios, taken together, for constraining the He
ionizing fraction and the ionizing spectrum. Domg\"{o}rgen \& Mathis
(1994) predict a high He ionizing fraction if the enhanced \NII\ 
emission is due to a dilute, hardened radiation field. A direct
measurement of the kinetic temperature is possible from a detection of
the \NII $\lambda 5755$ line (Osterbrock 1989). But these $V$ band lines
are expected to be an order of magnitude fainter than \Ha\ which puts
them at the limit of detectability (BTVS).

Finally, we summarize the steps that were taken to achieve the enhanced 
low ionization line ratios in Fig. 10. If \NIIr/\Ha\ $=$ 0.3 is
typical for a high $U$ model (e.g. HII region), a dilute radiation
field ($U<0.01$) in the neighbourhood of hot stars can double this
ratio (e.g.  Mathis 1986). Hardening the dilute radiation field with
interstellar absorption can produce line ratios closer to unity (e.g.
Sokolowski 1991). An additional 50\% increase is achieved with the
known gas phase depletions of refractory elements. Further enhancement
of this ratio (for a fixed abundance) requires that we selectively
heat the electrons without producing a higher ionization state of
nitrogen. 

\section{Conclusions}

We have succeeded in detecting ionized gas at and beyond the HI
cut-off radius in the nearby spiral galaxy, NGC 253. This galaxy is a
member of a small ensemble, the Sculptor group, with a low internal
dispersion and little or no associated hot medium. The detected
emission measures of the \Ha\ and \NII\ lines are sufficiently strong
that it is unlikely the source of the ionization is the metagalactic
UV background. The strength of the \NII\ line with respect to the \Ha\ 
line argues for an enhanced electron temperature at large galactic
radius compared with the inner HII regions. The dominant ionization
mechanism is suspected to be due to hot young stars in the inner regions 
which see the warped outer HI disk. We present a composite ionization 
model which may have some application to the HI edges of NGC 253, but
a full explanation requires additional heat sources. The kinematic
measurements confirm that the rotation curve is still rising at and
beyond the HI edge.  In some respects, since the HI disk ceases to be
detectable at only 1.2 \Redge, this is not an ideal object for testing
the original proposal of Bochkarev \& Sunyaev (1977).  But there are a
few objects which subtend a large solid angle on the sky and have HI
disks extending to beyond 2\Redge: these are the focus of subsequent
papers.

\acknowledgments JBH wishes to thank Oxford University for a Visiting
Fellowship and for their hospitality during the preparation of this
manuscript. We acknowledge extended dialogues with Dennis Sciama, Jon
Weisheit and James Binney.  We are indebted to Walter Dehnen for his
insights, and for carrying out a least-squares fit to our data. Both
an anonymous referee and the Scientific Editor, G.D. Bothun, made
substantive comments which improved the presentation of this
manuscript.  Keith Taylor and Sylvain Veilleux assisted in some of the
observations and the etalon was loaned to us by Brent Tully. We
acknowledge comments from George Field communicated by Dennis Sciama.
JBH wishes to thank Jim Sokolowski for permission to reproduce
unpublished work arising from extensive collaborations.

\newpage 
\references 
\reference Adams, F.C. \& Laughlin, G. 1996, ApJ, 468, 586
\reference Antonucci, R.R.J. \& Ulvestad, J.S. 1988, ApJ, 330, L97
\reference Arp, H. 1985, AJ, 90, 1012
\reference Bahcall, J.N. \etal\ 1991, ApJ, 377, L5
\reference Baldwin, J.\etal\ 1991, ApJ, 374, 580
\reference Beck, S.C. \& Beckwith, S.V. 1984, MNRAS, 207, 671
\reference Benjamin, R.A. \& Shapiro, P.R. 1993, In The Evolution of
Galaxies and their Environment, NASA Ames publ., p. 338
\reference Bica, E. \& Alloin, D. 1987, A\&A, 70, 281
\reference Bland-Hawthorn, J. 1997, PASA, in press
\reference Bland-Hawthorn, J., Ekers, R.D., van Breugel, W., Koekemoer, A. \& Taylor, K. 1995, ApJ, 447, L77
\reference Bland-Hawthorn, J. \& Maloney, P.R. 1997, ApJ, in press
\reference Bland-Hawthorn, J., Taylor, K., Veilleux, S. \& Shopbell, P.L. 1994, ApJ, 437, L95 
\reference Bochkarev, N.G. \& Sunyaev, R.A. 1977, Soviet Astr. 21, 542
(originally 1975, AZh, 54, 957) 
\reference Boggs, P.T., Byrd, R.H. \&
Schnabel, R.B. 1987, SIAM J. Sci. \& Stat. Computing, 8(6), 1052
\reference Bosma, A. 1978, University of Groningen thesis.  
\reference Bowyer, S. 1991, ARAA, 29, 59 
\reference Carignan, C. 1985, ApJ, 299, 59 
\reference Carignan, C. 1991, Proc.
Workshop on Dark Matter, Space Telescope Sci. Inst.
\reference Carignan, C. \& Freeman, K.C. 1985, 294, 494 
\reference Chabrier, G., Segretain, L. \& M'era, D.
1996, ApJ, 468, L21 
\reference Charlot, S. \& Silk, J. 1995, ApJ, 445, 124 
\reference Corbelli, E. \& Salpeter, E.E. 1988, ApJ, 326, 551 
\reference Corbelli, E., Schneider, S.E. \& Salpeter, E.E. 1989, AJ, 97, 390 
\reference Cowie, L.L. \& Songaila, A. 1986, ARAA, 24, 499
\reference Cruddace, R.\etal\ 1974, ApJ, 187, 497 \reference Dehnen,
W. \& Binney, J. 1997, MNRAS, in press \reference Diaz, A.I. 1989, in
Evolutionary Phenomena in Galaxies, eds. J.E. Beckman \& B.E.J. Pagel
(Cambridge: Cambridge University Press), 377 \reference Domg\"{o}gen,
H. \& Mathis, J.S. 1994, ApJ, 428, 647 \reference Dopita, M.A. \&
Sutherland, R. 1996, ApJS, 102, 161 \reference Dove, J.B. \& Shull,
J.M. 1994, ApJ, 423, 196 \reference Draine, B.T. \& Lee, H.M. 1984,
ApJ, 285, 89 \reference Evans, N. \& Dopita, M.A. 1985, ApJS, 58, 125
\reference Fadda\etal\ 1996, ApJ, 473, 670 
\reference Ferland, G.
\& Mushotsky, R. 1984, ApJ, 286, 42 \reference Ferland, G. \& Netzer,
H. 1983, ApJ, 264, 105 \reference Ferland, G. 1991, Cloudy code, OSU
Internal report, 91-01 \reference Ferland, G. 1992, ApJ, 389, L63
\reference Freeman, K.C. 1993, in Physics of Nearby Galaxies: Nature
or Nurture, ed Trinh Thuan \etal\ (Paris: Editions Frontieres), p 201.
\reference Freeman, K.C. 1996, in Unsolved Problems of the Milky Way,
ed L. Blitz and P. Teuben (Dordrecht: Kluwer), p 645.  \reference
Grevesse, N. \& Anders, E. 1989, In Cosmic Abundances of Matter, (New
York: AIP), p. 1
\reference Haardt, F. \& Madau, P. 1996, ApJ, 461, 20
\reference Halpern, J.P. \& Steiner, J.E. 1983, ApJ,
269, L37 \reference Haynes, M.P. \& Roberts, M.S. 1979, ApJ, 227, 767
\reference Henyey, L.G. \& Greenstein, J.L. 1941, ApJ, 93, 70 
\reference Henry, R.C. 1991, ARAA, 29, 89 
\reference Hogan, C.J. \& Weymann, R.J. 1987,
MNRAS, 225, 1P \reference Jenkins, E.B. 1987, in Interstellar
Processes, eds. D.J. Hollenbach \& H.A. Thronson (Dordrecht: Reidel),
533 
\reference Jura, M. 1980, ApJ, 241, 965
\reference Kawaler, S. 1996, ApJ, 467, L61 \reference Kamphuis, J.
1993, PhD, Groningen \reference Kondratyev, K. Ya. 1969, in Radiation
in the Atmosphere, Academic Press, New York \& London \reference
Koribalski, B., Whiteoak, J.B. \& Houghton, S. 1995, PASA, 12, 20
\reference Kurucz, R.L. 1979, ApJS, 40, 1 \reference Lee, Y.-W. 1993,
In The Globular Cluster-Galaxy Connection, eds. G.H. Smith \& J.P.
Brodie, ASP. conf. ser. 48, 142 \reference Liebert, J., Dahn, C.C.  \&
Monet, D.G. 1988, ApJ, 332, 891 \reference Lyon, J.G. 1975, ApJ, 201,
168 \reference Bania, T.M. \& Lyon, J.G. 1980, ApJ, 239, 173
\reference Maeder, A. 1990, A\&AS, 84, 139 \reference Malin, D.F.
1981, J. Phot. Sci., 29, 199 \reference Malin, D.F. 1983, In Astronomy
with Schmidt-type Telescopes, ed. M Capaccioli, IAU Coll. 78, 57
(Reidel: Dordrecht) \reference Maloney, P. 1993, ApJ, 414, 41
\reference Martin, P.G. \& Rouleau, F. 1990, preprint
\reference Mathewson, D.S. \& Ford, V.L. 1984, In Structure \& Evolution of the Magellanic
Clouds (Reidel: Dordrecht), IAU Symp. 108, 125 
\reference Mathewson, D.S., Cleary, M.N. \&
Murray, J.D. 1975, ApJ, 195, L97 
\reference Mathis, J. 1986, ApJ, 301, 423 
\reference Mathis, J., Rumpl, W. \& Nordsieck, K.H. 1977, ApJ, 217, 425 
\reference Mihalas, D. 1972, Non-LTE Model Atmospheres for B \& O stars (NCAR-TN/STR-76) 
\reference Miller, G.E. \& Scalo, J.M. 1979, ApJS, 41, 513 
\reference Molla, M., Ferrini, F. \& Diaz, A.I. 1996, ApJ, 466, 668
\reference Nordgren, T.E., Cordes, J.M. \& Terzian, Y. 1992, AJ, 104, 1465 
\reference Olling, R.P. 1995, 110, 591
\reference Osterbrock, D.E.  \& Martel, A. 1992, PASP, 104, 76
\reference Osterbrock, D.E. 1989, Astrophysics of Gaseous Nebulae \&
Active Galactic Nuclei, Univ. Sci. Books 
\reference Pagel, B.E.J. 1989, in Evolutionary Phenomena in Galaxies, eds. J.E. Beckman \& B.E.J. Pagel (Cambridge: Cambridge University Press), 201
\reference Puche, D. \& Carignan, C.  1991, ApJ, 378, 487 
\reference Puche, D., Carignan, C. \& van Gorkom, J.  1991, AJ, 101, 456 
\reference Rand, R. 1997, ApJ, 474, 129
\reference Raymond, J. 1990, ApJ, 365, 387
\reference Reynolds, R.J. \& Cox, D.P. 1992, ApJ, 400, L33
\reference Rots, A.H. 1975, A\&A, 45, 43
\reference Ryder, S.\etal\ 1996, PASA, 14, 81
\reference Ryu, D., Olive, K.A. \& Silk, J. 1990, ApJ, 353, 81 
\reference Salucci, P. \& Frenck, C.S. 1989, MNRAS, 237, 247
\reference Salpeter, E.E. 1955, ApJ, 121, 161
\reference Sargent, W.L.W. \& Steidel, C.C. 1990, in Baryonic Dark Matter, eds. D. Lynden-Bell \& G. Gilmore (Dordrecht: Kluwer), 223
\reference Savage, B.D. \& Massa, D. 1987, ApJ, 314, 380
\reference Schommer, R. \etal\ 1993, AJ, 105, 97
\reference Sciama, D.W. 1995, MNRAS, 276, L1 
\reference Sciama, D.W. 1996, Modern Cosmology \& The Dark Matter
Problem (Cambridge: Cambridge Univ. Press)
\reference Shields, J.C. 1992, ApJ, 339, L27
\reference Slavin, J.D., Shull, J.M. \& Begelman, M.C. 1993, ApJ, 407,
83
\reference Smith, G.P. 1963, Bull. Astron. Inst. Neth., 17, 203
\reference Sokolowski, J. 1994, preprint (unpubl.)
\reference Sokolowski, J. \& Bland-Hawthorn, J. 1991, PASP, 103, 911
\reference Spyromilio, J.  1995, MNRAS, 277, L59
\reference Staveley-Smith, L.\etal\ 1990, ApJ, 364, 23 
\reference Sutherland, R.S. \& Dopita, M.A. 1993, ApJS, 88, 253
\reference Tarter, C.B., Tucker, W.H. \& Salpeter, E.E. 1969, ApJ, 156, 943
\reference van Albada, T.S., Bahcall, J.N., Begeman, K. \& Sancisi,
R. 1985, ApJ, 295, 305 
\reference van Gorkom, J. 1993, In Environment \& Evolution of Galaxies, eds. 
J.M. Shull \& H.A. Thronson (Kluwer: Dordrecht), p. 345
\reference Veilleux, S. 1988, PhD thesis, U. C. Santa Cruz
\reference Vogel, S.N., Weymann, R., Rauch, M. \& Hamilton, T. 1995, ApJ, 441, 162
\reference Wakker, B.P.  1991, AA\&AS, 90, 495
\reference Wood, P. 1990, JRASC, 84, 150
\reference Yelle, R.V. \& Roesler, F. 1985, JGR, 90, 7568
\reference Zaritsky, D., Smith, R., Frenk, C., \& White, S.D.M. 1997, ApJ, 478, 39

\newpage
\noindent {\bf FIGURE CAPTIONS}
\medskip

\noindent {\bf Fig. 1.}
\label{B_four}
TAURUS-2 5\arcmin\ fields (SW1, SW2, SW3, SW4) superimposed on a $B$ 
image of NGC 253.  The
blue \NII\ line occurs at smaller galaxian radii compared with the
H$\alpha$ line in the same field.  There are four distinct fields with
two emission lines occurring in each. At the central position, the
etalon was used at two different spacings which meant that we obtained
line detections at four discrete positions.

\noindent {\bf Fig. 2.}
\label{HI_four}
Same as Fig. 1 except that the TAURUS-2 fields have been superimposed
onto a deep ATNF HI map (Koribalski, Whiteoak \& Houghton 1995). The
outermost HI contour corresponds to a column density of $4\times
10^{18}$ cm$^{-2}$.

\noindent {\bf Fig. 3.}
\label{zoom}
A magnified image of part of the SW3 field in Figs. 1 and 2. The field 
of view is
rotated so that the vertical axis lies parallel to the galaxy major
axis.  Clumpy \NII\ and H$\alpha$ emission is clearly visible. What is
not easily rendered is the diffuse emission between and at large
off-axis angles to the major axis.  Notice the extremely faint galaxy
continuum ($B\sim 25$ mag arcsec$^{-2}$) over much of the field. The
bright arcs are atmospheric OH lines.

\noindent {\bf Fig. 4.}
\label{spec1}
The emission-line spectrum at the HI edge (Field SW3 in Figs. 1 and 2) 
compared with
the off-field spectrum. The difference of these spectra is shown
below. Remarkably, the \NIIb\ line has a surface
brightness comparable to the H$\alpha$ surface brightness, as compared
with solar-abundance HII regions where the ratio is an order of
magnitude smaller. This result holds after removing the dense knots in
Fig. 3. The azimuthally averaged galaxy continuum underlies the
spectrum and corresponds to roughly $\mu_B =$ 23 mag arcsec$^{-2}$
below \NII\ falling to 25 mag arcsec$^{-2}$ below H$\alpha$. The
equivalent widths of the lines are 0.60AA\ and 1.8\AA\ respectively.

\noindent {\bf Fig. 5.}
\label{spec2}
The emission-line spectrum for the field beyond the HI edge (Field SW4 in 
Figs. 1 and 2). As we did not get a matching sky exposure, the spectra are
shown from binning over two different opening angles.  The top
spectrum results from binning over 25\deg\ about the major axis from
the optical axis; the lower spectrum arises from binning over the
entire field.  The difference of these spectra is shown below. The
\NII\ emission is clearly seen in the top spectrum and, indeed, in a
high contrast image. This line is completely washed out in the lower
spectrum. The H$\alpha$ emission is just visible in the high contrast
image, and does not begin to appear until we sum over the entire
field. The difference has been taken after weighting the top spectrum
for its lower signal to noise ratio.  Again, we see a very faint
galaxy continuum spectrum which is roughly $\mu_B =$ 26 mag
arcsec$^{-2}$ after azimuthal binning below the \NII\ line dropping to
$\mu_B =$ 27 mag arcsec$^{-2}$ below the H$\alpha$ line. The
equivalent widths of the lines are 7.8\AA\ and 9.9\AA\ respectively.

\noindent {\bf Fig. 6.}
\label{v_observed}
Kinematic measurements along the major axis of NGC 253 deduced from
VLA HI, TAURUS-2 \Ha\ and \NII\ lines. In $a$, the original rotation
curve of PCvG has been projected onto the sky with their inferred
orbit inclination at each radius. Three sets of measurements are
shown: these represent the approaching side (open circles), receding
side (squares), and from fitting to the full velocity field (filled
circles). In $b$, we show PCvG's original curves where the data have
been deprojected with the measured inclination at each radius. In $c$,
the solid line is the same curve as the squares in $a$. The filled
squares are the TAURUS-2 \NII\ measurements; the circles are the \Ha\ 
measurements.

\noindent {\bf Fig. 7.}
\label{v_model}
Representative fits to the PCvG and TAURUS-2 data using a 3-component
mass model (disk, halo, gas). The contribution from the HI surface
density (dotted line) is the same in all models.  In $a$, we adopt an
exponential disk (short-dashed line) and choose an isothermal sphere
for the spherical halo (long-dashed line) using the numerical approach
developed by Carignan (1985). In $b$, we have used the disk-halo
models of Dehnen \& Binney (1997): the least-squares fit was carried
out by W. Dehnen. In $c$, we have truncated the halo at a radius of 10
kpc to demonstrate a possible explanation for the last measured point.

\noindent {\bf Fig. 8.}
\label{white_dwarf}
The predicted \Ha\ emission measure \Em\ versus $B$ surface brightness
for an isothermal population of white dwarfs as a function of surface
density. The solid lines are isotherms for white dwarf temperatures in
the range $10^4 - 2\times 10^5$ K. The dashed lines are isochoric
lines indicating variations in white dwarf surface density in steps of
0.5 in dex starting at 1 pc$^{-2}$ at the top. The arrow indicates the
average emission measure and disk surface brightness upper limit at
the edge of the HI disk.  The filled circles are the expected
contribution to the $B$ band flux and \Ha\ emission measure for the
white dwarf population in the solar neighbourhood (assuming that all
photons are absorbed).  The Galactic disk is assumed to be 600 pc
thick and the white dwarf radius is assumed to be 0.14 R$_\odot$.

\noindent {\bf Fig. 9.}
\label{sok1}
The lower panel shows the relative ionization fraction of
important neutrals and ions, where the front of the gas slab ($\NH =
0$) is ionized by a dilute radiation field ($U\sim 10^{-4}$). The
middle panel shows the temperature and mean electron density structure
within the slab. The top panel shows the normalized line emissivities
and illustrates the relative dependence on electron temperature,
electron density, and ionization fraction.

\noindent {\bf Fig. 10.}
\label{sok2}
The dependence of five important line diagnostics on the strength of
the radiation field and hydrogen column density. The solid lines are
radiation bounded models where all ionizing photons are soaked up by
the gas. Also shown are matter bounded models where the slab has been
truncated at 1, 2, 5, 10 and 20 $\times$ 10$^{18}$ cm$^{-2}$. The
observed line ratio \NIIb/\Ha\ is a factor of 2.98 smaller than \NIIr/\Ha\
(Veilleux 1988). 
Fig.  9 gives some appreciation for the distinct trends shown by different
lines (e.g.  \OII\ vs. \OIII).

\end{document}